\documentclass[twocolumn,aps,prb,showpacs,superscriptaddress,amsfonts,amssymb,amsmath]{revtex4}

\begin{document}
\draft
\title{Quantum inductance and negative electrochemical capacitance at
finite frequency}

\author{Jian Wang $^*$}
\address{Center of theoretical and computational physics and
Department of Physics, The University of Hong Kong,
Pokfulam Road, Hong Kong, China\\}
\author{Baigeng Wang}
\address{Department of Physics, Nanjing University,
Nanjing, China \\}
\author{Hong Guo}
\address{Department of Physics, McGill University,
Montreal, Quebec, Canada \\}

\begin{abstract}
{We report on theoretical investigations of frequency dependent quantum 
capacitance. It is found that at finite frequency a quantum capacitor can be 
characterized by a classical RLC circuit with three parameters: a static 
electrochemical capacitance, a charge relaxation resistance, and a quantum 
inductance. The quantum inductance is proportional to the characteristic 
time scale of electron dynamics and due to its existence, the time dependent 
current can accumulate a phase delay and becomes lagging behind the applied 
ac voltage, leading to a negative effective capacitance.}
\end{abstract}


\maketitle

Understanding dynamic conductance of quantum coherent conductors is
a very important problem in nanoelectronics theory. When two quantum
coherent conductors form a double plate ``quantum capacitor", its
dynamic conductance $G(\omega)$ is given by the frequency dependent
electrochemical capacitance\cite{luryi,buttiker1,smith} $C_\mu
(\omega)$, $G(\omega) = -i\omega C_\mu (\omega)$, here $\omega$ is
the frequency. At low frequency, $C_\mu (\omega)$ can be expanded in
frequency and at the linear order, it is described\cite{buttiker1}
by an equivalent classical circuit consisting of a static capacitor
$C_\mu$ in series with a ``charge relaxation resistor" $R_q$. For a
conductor having a single spin-resolved transmission channel, $R_q$
was predicted\cite{buttiker1} to be half the resistance quantum,
$R_q=1/2 G_o$ where $G_o \equiv h/e^2$. The factor $1/2$ in $R_q$ is
of quantum origin\cite{buttiker1,nigg}, and has recently been confirmed
experimentally\cite{gabelli}. In the experiment of Gabelli {\it
et.al.}\cite{gabelli}, a submicron 2DEG quantum dot (QD) is capacitively 
coupled to a gold plate forming a double plate capacitor, where the 
QD connects to the outside reservoir by a single channel quantum point 
contact (QPC). The dynamic conductance $G(\omega)$ is then measured at 1.2GHz, 
and the data is well fit to the equivalent circuit characterized by 
two parameters $(C_\mu, R_q)$.

The experiment of Gabelli {\it et.al.}\cite{gabelli} opened the door for
elucidating important and interesting physics of high frequency quantum
transport in meso- and nano-scale devices. An important question is what
happens to electrochemical capacitance at higher frequencies beyond the linear
$\omega$ regime, and in particular whether the two-parameter $(C_\mu, R_q)$
equivalent circuit is adequate at higher frequencies to describe a quantum
capacitor. It is the purpose of this paper to address these issues.

In the following, we report a microscopic theory for high frequency quantum
transport in a two-plate quantum capacitor. Our results show that to 
characterize its high frequency dynamic response, one needs---in addition 
to $C_\mu, R_q$, a new quantity $L_q$ having the dimension of inductance. 
$L_q$ is found to have purely quantum origin and will be named ``quantum 
inductance". Therefore, the frequency dependent electrochemical capacitance 
of a quantum capacitor $C_\mu(\omega)$ is equivalent to a classical RLC 
circuit characterized by three parameters $(C_\mu, R_q, L_q)$, at high 
frequency. Due to $L_q$, electrons dwell in the neighborhood of the capacitor 
plates causing a phase delay. At low frequencies, the dynamic response is 
capacitive-like and voltage lags current. At larger frequencies when 
$\omega > 1/\sqrt{C_\mu L_q}$, inductive behavior dominates and voltage leads 
current: in this case the quantum capacitor gives a negative capacitance 
value. It is, indeed, surprising that a quantum capacitor can give an 
inductive dynamic response. For the experimental setup of 
Ref.\onlinecite{gabelli}, we estimate that when $\omega \sim 3$GHz, the 
predicted high frequency effects should be observable.

Let's first work out a simple expression for the frequency 
dependent electrochemical capacitance $C_\mu(\omega)$ following the work
of Buttiker\cite{buttiker1}. We consider a two-plate
capacitor similar to the experiment of Gabelli {\it et.al.}\cite{gabelli}: 
a QD---labelled I, and a large metallic electrode---labelled II. Each plate 
is connected to the outside world through its lead and a time dependent bias 
$v_{1,2}$ is applied across the two leads. We consider small amplitudes of
$v_{1,2}$ so as to focus on the linear bias regime. Under the action of
such a bias, the two capacitor plates develop their own frequency dependent 
electric potential $U_{I,II}(\omega)$. The charge on plate-I is equal to the 
sum of the injected charge and induced charge: $Q_I=Q_I^{inj} + Q_I^{ind}$. 
In the linear regime, the injected charge is proportional to bias 
$v_1(\omega)$: $Q_I^{inj} = e^2 D_I(\omega) v_1(\omega)$ where $D_I(\omega)$
%is the average local density of states (DOS) of plate-I at frequency $\omega$. 
is the generalized global density of states (DOS) of plate-I at frequency 
$\omega$. 
The induced charge, on the other hand, is in general related to a nonlocal 
Lindhard function\cite{buttiker1} whose calculation is simplified by applying 
the Thomas-Fermi approximation. Within this approximation, the induced charge
is proportional to the induced potential $U_I$ on the plate: 
$Q_I^{ind}=-e^2 D_I(\omega)U_I$ where the minus sign indicates that the
charge is induced. Putting things together, we obtain
$Q_I=e^2D_I(\omega)(v_1-U_I)$. Clearly, the same charge $Q_I$ can be calculated
by the usual electrostatic geometric capacitance $C_o$: 
$Q_I=C_o(U_I - U_{II})$. We therefore obtain a relationship: 
$C_o(U_I - U_{II})=e^2D_I(\omega)(v_1-U_I)$. Applying the same argument to 
plate-II, we similarly obtain $-C_o(U_I - U_{II})=e^2D_{II}(\omega)
(v_2-U_{II})$. Finally, the same charge $Q_I$ can also be obtained from the 
definition of the electrochemical capacitance: $Q_I=C_\mu(\omega)(v_1 - v_2)$.
These three relations allow one to derive the following expression for
$C_\mu(\omega)$:
\begin{eqnarray}
\frac{e^2}{C_\mu(\omega)} = \frac{e^2}{C_o}+\frac{1}{D_I(\omega)} +
\frac{1}{D_{II}(\omega)} \label{final1}
\end{eqnarray}
This result resembles the one obtained by B\"uttiker for the static
capacitance\cite{buttiker1}. An important difference is that the 
frequency dependent electrochemical capacitance in Eq.(\ref{final1}) is a 
complex quantity: its real part is a measure to the electrochemical 
capacitance and its imaginary part is proportional to the frequency dependent 
charge relaxation resistance.

To proceed further, we need to calculate the frequency dependent DOS
$D_{I,II}(\omega)$. Following Ref.\onlinecite{nigg}, the generalized local
DOS of plate-$\alpha$ of the capacitor can be expressed in terms of Green's functions:
\begin{equation}
\frac{dn_\alpha (\omega)}{dE}=\int \frac{dE}{2\pi } 
\frac{f-{\bar f}}{\hbar\omega}
[{\bar G}^{r}\Gamma _{\alpha}G^{a}]_{xx} \label{dos}
\end{equation}
where the subscript $x$ labels space coordinates and $D_\alpha(\omega)=
Tr[dn_\alpha(\omega)/dE]$. In Eq.(\ref{dos}),  $f$ is the Fermi function and 
$\bar{f}\equiv f(E_+)$ with $E_+\equiv E+\hbar\omega$; $G^{r,a}=G^{r,a}(E)$ 
is the retarded/advanced Green's function at energy $E$ and 
$\bar{G}^r \equiv G^r(E+\hbar\omega)$; $\Gamma_\alpha$ is the linewidth 
function 
describing the coupling strength between plate-$\alpha$ with its lead. These 
quantities can be calculated in straightforward manner when the Hamiltonian 
of the capacitor model is specified\cite{ma1,wbg}.

For the quantum capacitor of Ref.\onlinecite{gabelli}, plate-I is a
QD and plate-II is a metal gate. Since the metal gate has much greater DOS, 
{\it i.e.} $D_{II} >> D_I$, we can safely neglect the $D_{II}$ term in 
Eq.(\ref{final1}). For a QD with one energy level $E_0$ and connected to one 
lead, its Green's function $G^r = 1/(E-E_0+i\Gamma_L/2)$ where $\Gamma_L$ is 
the linewidth function of the lead. With this Green's function, the frequency 
dependent DOS can be easily calculated from Eq.(\ref{dos}), we obtain:
\begin{eqnarray}
&&D_I(\omega) = \frac{\Gamma_L}{2\pi\hbar\omega(\hbar\omega
+i\Gamma_L)}[\frac{1}{2}\ln\frac{\Delta^2} {\Delta_+ \Delta_-}
\nonumber \\
&&-i(\arctan \frac{\Delta E-\hbar\omega}{\Gamma_L/2} - \arctan
\frac{\Delta E+\hbar\omega}{\Gamma_L/2})] \label{dos2}
\end{eqnarray}
where $\Delta=\Delta E^2+\Gamma_L^2/4$, $\Delta_\pm=(\Delta E \pm
\hbar\omega)^2+\Gamma_L^2/4$, and $\Delta E=E_F-E_0$. At resonance
$\Delta E=0$, we obtain $Re(D_I(\omega))=
[-x\ln(4x^2+1)+2\arctan(2x)]/(2\pi \Gamma_L x)/(x^2+1)$ with
$x=\hbar\omega/\Gamma_L$. Hence $Re(D_I(\omega))$ is positive for
small $x$ and negative for large $x$, {\it i.e.} there is a sign
change. A similar behavior is also found for the system away from
the resonance. Due to this sign change of $Re(D_I)$, from Eq.(\ref{final1}) 
the frequency dependent electrochemical capacitance 
$C_R \equiv Re[C_\mu(\omega)]$ can become negative.

To be more specific, we fix the classical capacitance of QD $C_0=1$fF which 
is a typical value for QD with area of $\sim 1\mu m^2$. Fig.1 plots 
$C_R=Re[C_\mu(\omega)]$ (real part) and $C_I=Im[C_\mu(\omega)]$ 
(imaginary part) versus frequency for different values of $\Gamma_L$, by 
setting $\Delta E$ and temperature to zero. We observe that $C_R$ is positive 
at small frequency and becomes negative at larger frequency. For
instance, $C_R$ becomes negative at a ``critical" frequency
$\omega_c\sim 10GHz$ for $\Gamma_L=10\mu eV$. This critical
frequency can be smaller for smaller linewidth function $\Gamma_L$.
We note that it is not difficult to achieve $\Gamma_L=10\mu eV$
experimentally: in Ref.\onlinecite{fujisawa}, $\Gamma_L$ between
$1\mu eV $ to $5 \mu eV$ has been realized. As will be discussed
below, the effective $\Gamma_L$ in the experiment of
Ref.\onlinecite{gabelli} is tunable by a gate voltage so that the
critical frequency at which the negative capacitance occurs can be
even smaller. The inset of Fig.1 also shows that as we increase $\omega$, the
imaginary part of $C_\mu(\omega)$ starts from zero, reaches a peak
value around $\omega_c$ and then decays to zero. The negative
capacitance at large frequency can be understood as follows. For a
classical capacitor, a charge is accumulated across the capacitor
induced by an external voltage. The current and voltage has a fixed
phase relationship: the voltage lags behind current with a phase
$\pi/2$. For a quantum capacitor at low frequency, there exists a
charge relaxation resistance $R_q=h/(2e^2)$ for a single channel
plate, therefore the charge build-up time is the RC-time $\tau_{RC}
= R_q C_\mu$. For $C_\mu=1$fF and $R_q=h/(2e^2)$, this RC-time is
about $\tau_{RC}=13$ps. If the external voltage reverses sign, the
charge accumulation will follow the voltage and also reverse sign in
due time. When frequency is low, namely when $\omega <<1/\tau_{RC}=77GHz$, 
the charge build up follows the ac bias almost instantaneously just like
a classical capacitor, thus the capacitance is positive. When
frequency is high, there is a phase difference between the ac bias
and the charge build up. For frequencies comparable to $1/\tau_{RC}$, the 
charge build up can not follow the ac bias, thereby the capacitance may be 
negative.

While the above argument explains why it is possible to have negative 
capacitance, it would indicate a critical frequency to be near $77$GHz. Our 
results (Fig.1) show that the calculated $\omega_c$ is actually much smaller. 
This is because there exists a second relevant time scale in the QD, 
{\it i.e.} the dwell time $\tau_d$ which is the time spent by electrons 
inside the QD. The dwell time $\tau_d$ can be calculated for specific 
systems\cite{gasparian}. Importantly, at resonance the electrons can dwell 
inside the quantum dot for a long time. For instance, for our QD with
$\Gamma_L=10\mu$eV,
we found $\tau_d=260$ps while $\tau_{RC}=12$ps (since $C_\mu=0.9fF$). In other words, 
$\tau_d >> \tau_{RC}$. Such a $\tau_d$ corresponds to a frequency of $4$GHz,
much less than $1/\tau_{RC}$. In other words, when ac frequency is
greater than $1/\tau_d$, the charges dwell inside the quantum dot and
cannot respond to the ac voltage change. As a result, current becomes
lagging behind voltage, leading to a negative capacitance. This picture agrees 
very well with the numerical results (Fig.1).

Having determined the general behavior of $C_\mu(\omega)$ for the quantum 
capacitor, in the following we determine how to simulate this quantity using 
a classical circuit. Expanding Eq.(\ref{final1}) into a Taylor series to 
second order in $\omega$ with the help of Eq.(\ref{dos2}) at resonance, we 
obtain:
\begin{equation}
C_\mu(\omega) = C_\mu + i\omega C_\mu^2 \frac{h}{2e^2} - \omega^2
C_\mu^3 \frac{h^2}{4e^4} +\omega^2 C_\mu^2 \frac{ h^2}{12\pi
\Gamma_L e^2} \label{cla1}
\end{equation}
where $C_\mu=C_\mu(0)$ on the right hand side is the static electrochemical 
capacitance. This result is equivalent to that of a classical RLC circuit, 
as follows. For a classical RLC circuit with capacitance $C_\mu$, resistance 
$R_q$ and inductance $L_q$, the dynamic conductance is
\begin{equation}
G(\omega) = -i\omega C_\mu/(1-\omega^2L_q C_\mu-i\omega C_\mu R_q)\
\ . \label{circuit1}
\end{equation}
Expanding this expression in power series of $\omega$, we obtain
\begin{equation}
G(\omega) = -i\omega C_\mu + \omega^2 C_\mu^2 R_q +i\omega^3 C_\mu^3
R_q^2 -i\omega^3 C_\mu^2 L_q\ . \label{circuit2}
\end{equation}
Because for a capacitor $G(\omega)=-i\omega C_\mu(\omega)$, we obtain the 
result that our quantum capacitance Eq.(\ref{cla1}) is equivalent to the 
classical RLC circuit model of Eq.(\ref{circuit2}). Comparing these two 
equations we readily identify $R_q = h/(2e^2)$---a result first obtained by 
B\"uttiker and co-workers\cite{buttiker1}. Importantly, a new quantity---the
equivalent inductance, is identified as $L_q=h^2/(12\pi e^2 \Gamma_L)$. In 
terms of dwell time $\tau_d$ and charge relaxation resistance $R_q$, we obtain:
\begin{equation}
L_q=R_q \tau_d/12
\label{induc}
\end{equation}
where $\tau_d=4\hbar/\Gamma_L$.

What is the reason that a quantum capacitor at finite frequency needs to be 
modeled by a classical RLC circuit (instead of a RC circuit)? This is due to 
the role played by the large dwell time $\tau_d$ of the QD. When electrons 
dwells a long time $\tau_d$ inside the QD, the interaction between electrons 
become an important piece of physics which, in our theory, is modeled by the 
induced self-consistent potential $U_{I,II}$ discussed above. Such an
interaction gives rise to the physics of induction, and resulting to
the quantity $L_q$ of Eq.(\ref{induc}). Indeed, the explicit dependence on 
$\tau_d$ by $L_q$ also confirms the important role played by the dwell time. 
Because $L_q$ is determined by $\tau_d$ as well as fundamental constants $h$ 
and $e$, it is of purely quantum origin and can be called quantum inductance.

Fig.2 compares the fitting of classical RLC circuit
Eq.(\ref{circuit1}), with the full quantum result of Eq.(\ref{final1}). 
They compare very well for the entire range of the frequency---if we treat 
$R_q$ as a function of $\omega$. Indeed, while $R_q$ has so far been a constant
$h/(2e^2)$ as identified through the Taylor expanded 
Eqs.(\ref{cla1},\ref{circuit2}), it is actually a function of $\omega$ by the 
more general expression Eq.(\ref{circuit1}). The inset of Fig.3 plots the 
general $R_q=R_q(\omega)$ obtained numerically, and we observe it to be a 
slowly increasing function of $\omega$. As expected, in the small frequency 
limit, $R_q(\omega)$ recovers the result of half resistance quantum. For 
$\Gamma_L=50\mu eV$, $R_q(\omega)$ deviates from $h/2e^2$ at about $5 GHz$. 
%For $\Gamma_L=10\mu eV$, the deviation starts at $\omega=1GHz$. 
We have also attempted using three {\it constant} parameters $C_\mu$, $L_q$ and 
$R_q=h/(2e^2)$ into Eq.(\ref{circuit1}) to compare with the full quantum 
result of Eq.(\ref{final1}), a reasonable agreement is obtained (inset of Fig.2) 
although not as good as that shown in Fig.2.

The situation is somewhat different for quantum inductance $L_q$ when the 
system is \emph{off} resonance ($\Delta E \neq 0$ in Eq.(\ref{dos2})). In 
this case the dwell time $\tau_d$ becomes too small to be relevant and another 
time scale becomes important, namely the tunneling time $\tau_t$ for 
electrons to go in/out of the QD. The further away $E$ is from $E_0$, the 
longer is $\tau_t$. Hence in Eq.(\ref{induc}) $\tau_d$ should be replaced 
by $\tau_t$ for off resonance. Our result shows that the fitting of full 
quantum capacitance $C_\mu(\omega)$ using classical parameters $C_\mu$, 
$L_q$ and $R_q(\omega)$ is still quite good for off resonance. This further 
supports the conclusion that the frequency dependent quantum capacitance can 
be described by a classical RLC circuit with static electrochemical 
capacitance, charge relaxation resistance, and a quantum inductance.

Finally, we perform a numerical calculation of the dynamic conductance for 
the device structure of Ref.\onlinecite{gabelli}. In terms of scattering 
matrix, the DOS of Eq.(\ref{dos}) for a capacitor can be re-written as\cite{ma1,nigg}
\begin{equation}
D_I(\omega)=i\int \frac{dE}{2\pi }
\frac{f-{\bar f}}{\hbar^2\omega^2} [1-s^\dagger_{LL}(E_+) s_{LL}(E)]
\end{equation}
with\cite{gabelli} $s^\dagger_{LL}(E)=(r-e^{i\phi})/(1-r e^{i\phi})$, 
$\phi=2\pi E/\Delta$, $r^2=1-T_{QPC}$ and $T_{QPC}=1/[1+\exp(-(V_g+V_0)/\Delta V_0)]$ 
which is the transmission coefficient of the QPC in the experimental
setup\cite{gabelli}. Fig.3 shows the dynamic conductance $G(\omega)$
vs gate voltage $V_g$ using our theory presented above.
When $\omega=1.2GHz$ (open circle), our results agrees very well\cite{foot3}
with the experimental data of Ref.\onlinecite{gabelli}. When
$\omega=3GHz$ (open square), our theory predicts that the imaginary
part of $G(\omega)$ which is the electrochemical capacitance, goes
to negative. For even larger frequency $\omega=5GHz$ the effect is
more significant. To understand why one can observe a negative
capacitance at small frequency such as $3GHz$, we note that since
electron entering the QD  has to first pass the QPC\cite{gabelli}: this 
QPC serves as a barrier (with an effective barrier height $1/\Gamma$) that is
controlled by the gate voltage. At small gate voltage, $T_{QPC}$ is nearly zero 
and goes to one at large $V_g$. Hence the effective $\Gamma$ for small gate 
voltage is quite large making $\omega_c$ much smaller. Since the experiment of
Ref.\onlinecite{gabelli} is performed at $\omega=1.2$GHz, we assume
that it is not too difficult to push the frequency to $3$GHz so that
the effect of quantum inductance can be observed experimentally.
Indeed, a single-wall carbon nanotube transistor operated at
$2.6$GHz has been demonstrated\cite{Li} and measurement of current
fluctuation at frequency from 5 to 90 GHz has been
reported\cite{Deblock}.

In summary, we found that at finite frequency, a quantum capacitor
consisting of a quantum dot and a large metal conductor is
equivalent to a classical RLC circuit with three basic parameters: a
static electrochemical capacitance $C_\mu$, a charge relaxation
resistance $R_q$, and a quantum inductance $L_q$. It is found that
$L_q \sim R_q \tau$ where $\tau$ is the characteristic time scale
for the quantum dot such as the dwell time $\tau_d$ or the tunneling
time $\tau_t$. Due to the phase delay by the quantum inductance, the
dynamic current can lag behind of the applied ac voltage, giving
rise to a negative capacitance. Our numerical results show that this
effect should be detectable experimentally using the present device
technology.

{\bf Acknowledgments.} We thank Prof. K. Xia for useful discussions. This work
is supported by a RGC grant (HKU 7048/06P) from the HKSAR and LuXin Energy 
Group (J.W); NSF-China grant number 10474034 and 60390070 (B.G.W); NSERC 
of Canada, FQRNT of Qu\'{e}bec and Canadian Institute of Advanced Research (H.G). 

\bigskip
\noindent{$^{*)}$ Electronic address: jianwang@hkusub.hku.hk}

%Fig.1
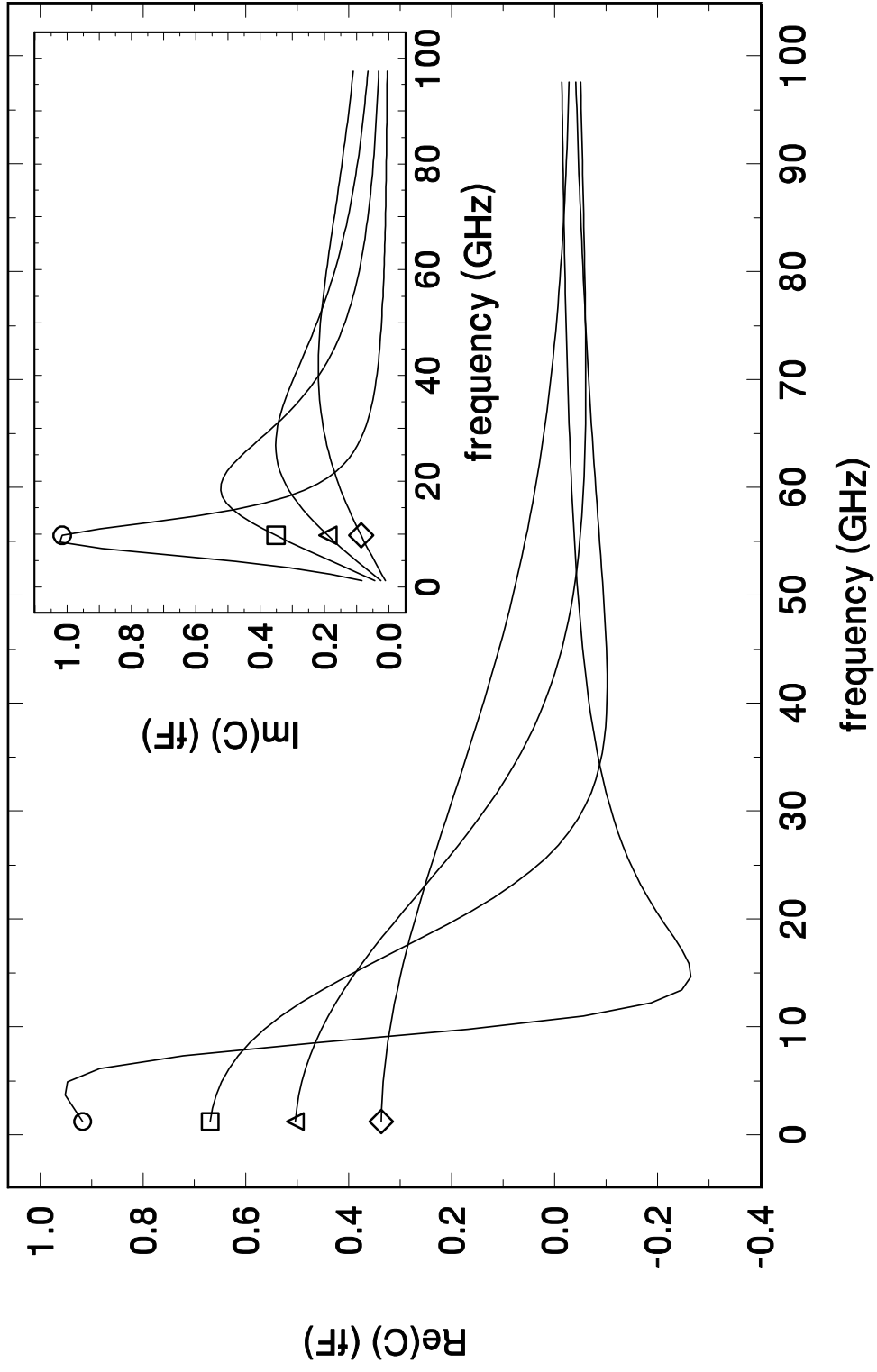
\begin{figure}[tbp]
\caption{
Frequency dependent capacitance $C_\mu(\omega)$ versus frequency. Main
figure is for $C_R$ and inset is for $C_I$.
Here $\Gamma_L=10\mu eV$ (for the curve with the open circle), 
$50 \mu eV$ (open square), $100 \mu eV$ (open triangle), 
and $200 \mu eV$ (open diamond). Here $C_o=1fF$.
}
\label{fig1}
\end{figure}

%Fig.2
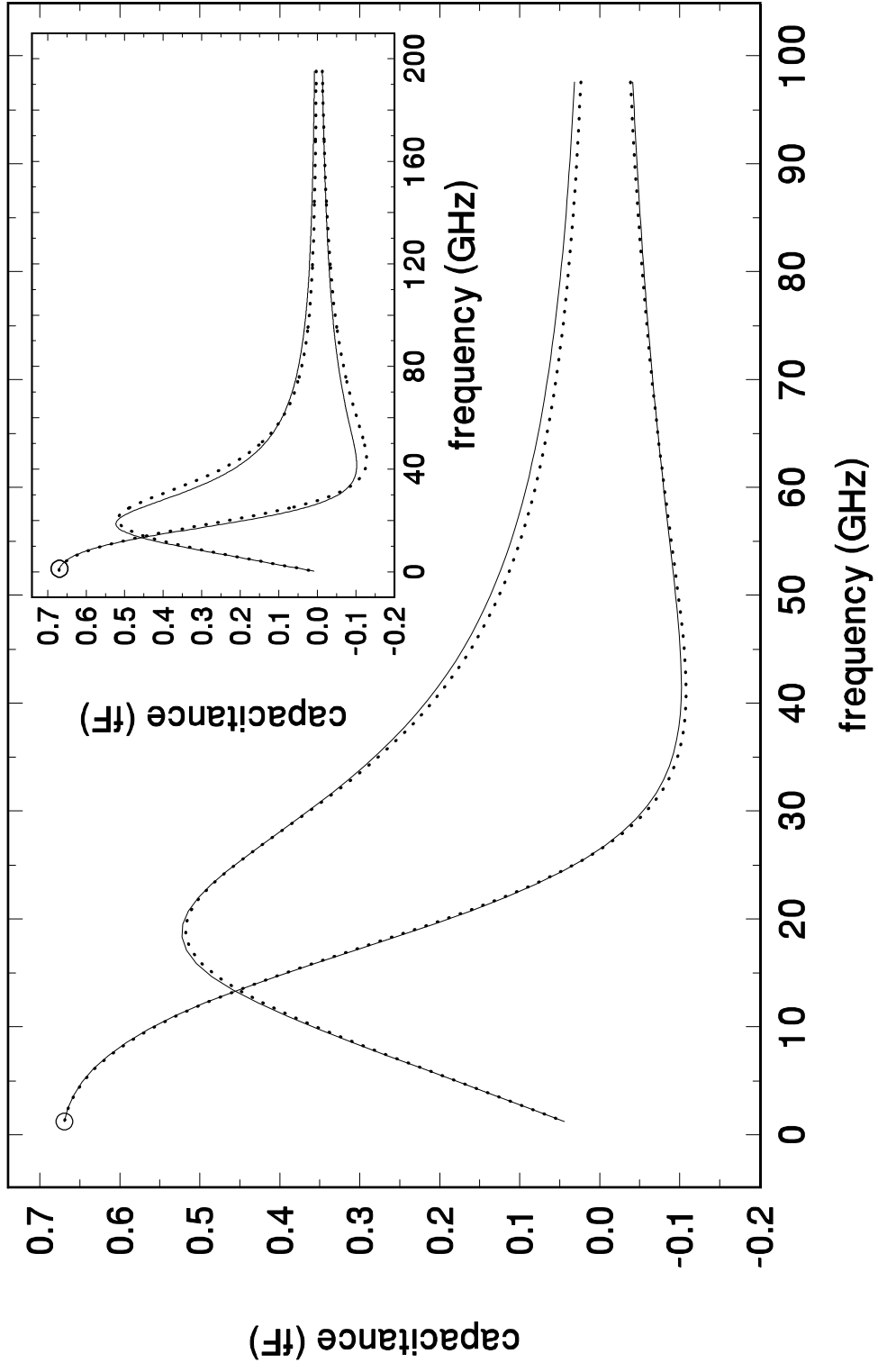
\begin{figure}[tbp]
\caption{Comparison of the full quantum capacitance $C_\mu(\omega)$ (solid line)
to that obtained by the classical RLC circuit model (dotted line) for $\Gamma_L=50\mu eV$. 
Here $Re[C_\mu(\omega)]$ is indicated by open circle. Inset: similar fit but
using three constant parameters.
}
\label{fig2}
\end{figure}

%Fig.3
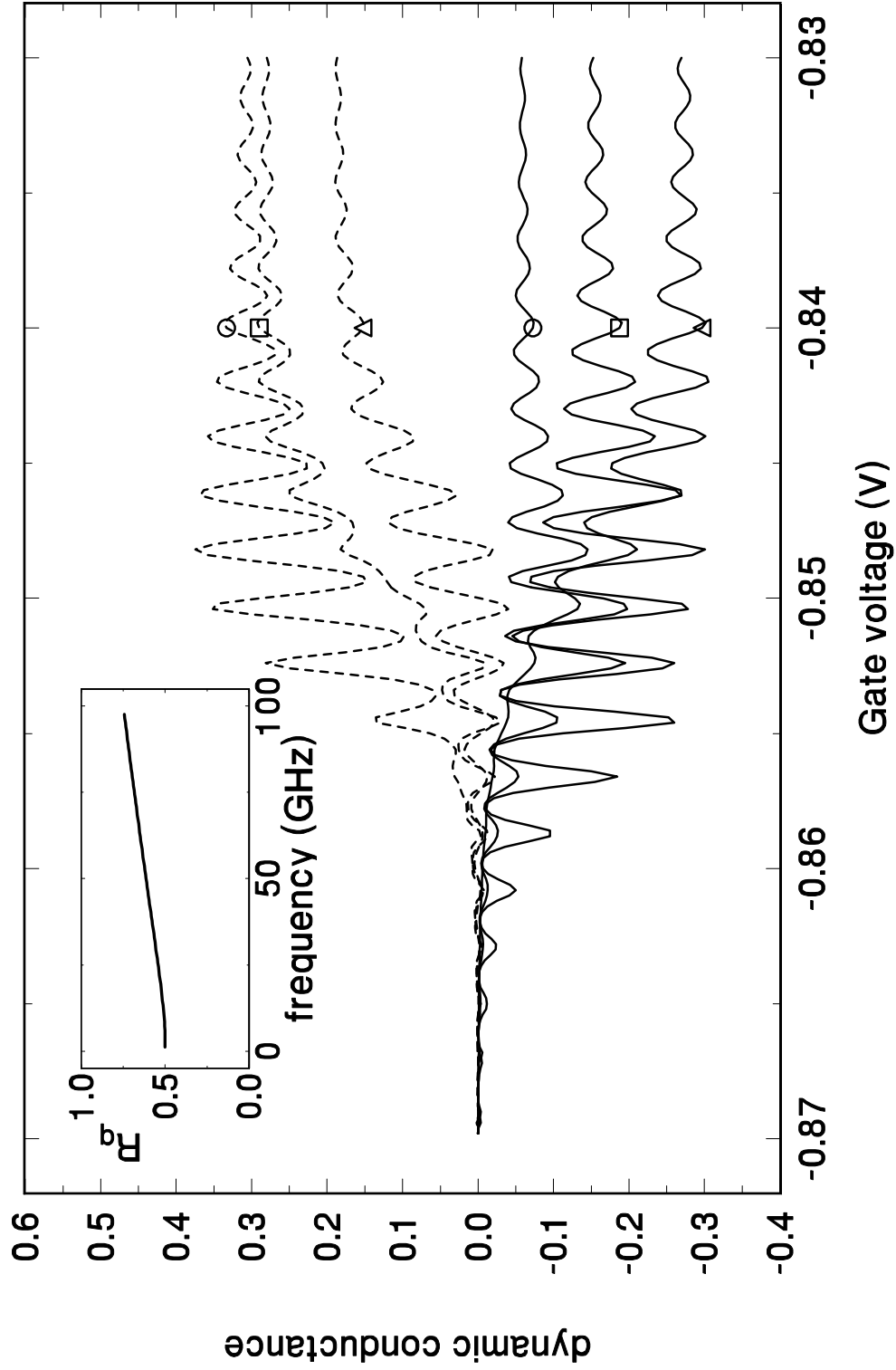
\begin{figure}[tbp]
\caption{The dynamic conductance $G(\omega)$ (unit $e^2/h$) versus gate 
voltage at different frequencies $\omega=1.2, 3, 5 GHz$. Solid line 
$-Re[G]/\omega$, dotted line $-Im[G]/\omega$. Other parameters: $\Delta=500mK$, 
$\alpha=3000$, $V_0=-0.85$, $\Delta V_0=0.003$, $C_0=4fF$, and temperature 
$50mK$. For purpose of illustration, we divided $G(\omega)$ by $\omega$. Inset:
frequency dependent $R_q$ (unit $h/e^2$) vs frequency.
}
\label{fig3}
\end{figure}

\begin{thebibliography}{00}
\bibitem{foot3} 
In classical RC circuit theory, the convention is writing 
$v(t)=v_0 \exp(i\omega t)$, while in quantum transport theory the convention 
is $v(t)=v_0 \exp(-i\omega t)$. This is responsible for the sign difference 
between our Fig.3 and the experimentally data of Ref.\onlinecite{gabelli}.
\end{thebibliography}
\end{document}